\documentclass[reprint,amsmath,amssymb,aps,floatfix,groupedaddress,superscriptaddress,nofootinbib,preprintnumbers]{revtex4-1}
\usepackage{float}
\usepackage{amsmath}
\usepackage[pdftex]{hyperref}
\usepackage{graphicx,color}
\usepackage{dcolumn}
\usepackage{bm}
\usepackage{here}
\usepackage{comment}
\usepackage[paperwidth=210mm,paperheight=297mm,centering,hmargin=1.5cm,vmargin=2.0cm]{geometry}
\usepackage{tabularx}
\graphicspath{{./fig/}}

\allowdisplaybreaks[1]

\newcommand{\ene}{$\sqrt{s_{\rm{NN}}}=$~}
\newcommand{\enebes}{$\sqrt{s_{\rm{NN}}}=$ 7.7, 11.5, 14.5, 19.6, 27, 39, 62.4 and 200 GeV}

\makeatletter
\newsavebox{\@brx}
\newcommand{\llangle}[1][]{\savebox{\@brx}{\(\m@th{#1\langle}\)}%
  \mathopen{\copy\@brx\mkern2mu\kern-0.9\wd\@brx\usebox{\@brx}}}
\newcommand{\rrangle}[1][]{\savebox{\@brx}{\(\m@th{#1\rangle}\)}%
  \mathclose{\copy\@brx\mkern2mu\kern-0.9\wd\@brx\usebox{\@brx}}}
\makeatother

\begin{document}

\title{Volume fluctuation and multiplicity correlation on higher-order cumulants}

\author{Tetsuro Sugiura}
\email{tsugiura@rcf.rhic.bnl.gov}
\affiliation{Tomonaga\,Center\,for\,the\,History\,of\,the\,Universe,\,University\,of\,Tsukuba,\,Tsukuba,\,Ibaraki\,305,\,Japan}
\author{Toshihiro Nonaka}
\email{tnonaka@rcf.rhic.bnl.gov}
\affiliation{Key\,Laboratory\,of\,Quark\,\&\,Lepton\,Physics\,(MOE)\,and\,Institute\,of\,Particle\,Physics,\,Central\,China\,Normal\,University,\,Wuhan\,430079,\,China}
\author{ShinIchi Esumi}
\email{esumi.shinichi.gn@u.tsukuba.ac.jp}
\affiliation{Tomonaga\,Center\,for\,the\,History\,of\,the\,Universe,\,University\,of\,Tsukuba,\,Tsukuba,\,Ibaraki\,305,\,Japan}


\begin{abstract}

Initial volume fluctuation (VF) arising from the participant fluctuation would be the
background which should be subtracted experimentally from the measured higher-order cumulants. 
We study the validity of the Volume Fluctuation Correction (VFC) on higher-order net-proton cumulants 
by using simple toy model and UrQMD model in Au+Au collisions at \ene 200 GeV for various centrality definitions.
The results are compared to the conventional data driven method called Centrality Bin Width Correction (CBWC).
We find VFC works well in toy model assuming independent particle production (IPP), but does not seem to work well in UrQMD model.
It is also found that cumulants are strongly affected by the multiplicity correlation effect as well as the centrality resolution effect.
These results show that neither VFC nor CBWC are perfect method.
Thus, both methods should be compared in the real experiment.

\end{abstract}
\maketitle

\newcommand{\ave}[1]{\ensuremath{\langle#1\rangle} }

\section{Introduction}
\label{sec:intro}
In heavy-ion collision experiments, the study of event-by-event fluctuation is a powerful tool to characterize the thermodynamic properties of the hot and dense QCD matter.
According to the Lattice QCD calculations, an analytic crossover exists at $\mu_{B}\approx0$. 
However, the detailed structure of the QCD phase diagram at larger $\mu_{B}$ resion including the location of the critical point \cite{susceptibility,correlation,Asakawa:2009aj}, which is the end point of the first-order phase transition boundary, is not known well.  
According to the theoretical predictions, higher-order fluctuations of conserved quantities, such as net-baryon, net-charge and net-strangeness, diverge near the critical point.
Experimentally, net-proton and net-kaon are measured as a proxy for the net-baryon and net-strangeness.
In order to study the detailed structure of the QCD phase diagram, Beam Energy Scan I (BES-I) program was carried out from 2010 to 2014 at \enebes~at RHIC.
The STAR experiment published up to the fourth-order fluctuations of net-proton, net-charge and net-kaon distributions \cite{net_proton,net_charge,net_kaon}, and recently up to the sixth-order fluctuations of net-proton and net-charge distributions are measured \cite{NonakaQM}.
At fourth-order fluctuation of net-proton with transverse momentum 0.4$<p_{T}<$2 GeV/$c$, non-monotonic behaviour has been observed in 0-5\% centrality at low energy regions with large uncertainties \cite{Luo:2015doi}.
In order to shrink the errors observed in BES-I, Beam Energy Scan II (BES-II) program has been conducted to investigate lower energy regions with high statistics from 2019. 
The NA61/SHINE experiment at CERN and the HADES experiment at GSI also inventigated heavy-ion collisions at low energy regions in order to discover the diagram \cite{Abgrall:2014xwa,Szala:2018rrr}.

Initial volume fluctuation (VF) is the event-by-event fluctuation of number of participant nucleons ($N_{W}$) in heavy-ion collision experiment \cite{Gorenstein:2011vq,Skokov:2012ds}.
On higher-order event-by-event fluctuation analysis, the VF is one of the experimental backgrounds which should be taken into account.
Specifically, it is well known that experimentally measured cumulants of net-particles are artificially enhanced due to the VF \cite{Xiaofeng1}.
One possible way to suppress the VF is applying Centrality Bin Width Correction (CBWC) \cite{Xiaofeng1} which is a conventional data driven method, and
the STAR experiment has been using this method.
However, there might be some residual fractions of VF backgrounds even with CBWC.
Recently, a new correction method called Volume Fluctuation Correction (VFC) \cite{VFC} is proposed under the assumption of the independent particle production (IPP) model.
The HADES experiment applied VFC to the experimental data \cite{HolzmannQM}, and the validity of the VFC has already been studied by using simple toy model \cite{VFC}.
However, it is not obvious that VFC works well when we apply VFC to the experimental data because IPP model is expected to be broken in the real experiment.
Thus, it is important to study the validity of the VFC in more realistic situations without IPP assumption.

More importantly, it is expected that the value of cumulants strongly depends on the centrality definitions, which are often determined by experimentally measured charged particle multiplicities.
It is known that there are two effects called centrality resolution effect and auto-correlation effect \cite{Xiaofeng1}.
The centrality resolution effect depends on multiplicity scale, and related to the VF.
On the other hand, it is expected that auto-correlation can be removed by using different kinematic window between centrality definitions and cumulant measurements. 
However, even though we remove auto-correlation, multiplicity correlation between centrality and the measured cumulant may still exist, which will also affect the measured cumulants.
We call this effect multiplicity correlation effect.
The auto-correlation and the centrality resolution effects were studied for some cases \cite{Xiaofeng1,CBWCN}, but it is still not clear how cumulants are affected by the multiplicity correlation.
At STAR experiment, centralities are defined using charged particle multiplicities in mid-rapidity regions by the Time Projection Chamber (TPC) in BES-I.
From BES-II, the Event Plane Detector (EPD), whose acceptance is 2.1$<|\eta|<$5.1, has been used for the centrality determination \cite{EPD}, but it is unknown that how cumulants are affected when we use EPD instead of TPC.
In order to answer this question, detailed studies about both VF and multiplicity correlation effect are needed.

This paper is organized as follows.
In Sec.~\ref{sec:Observables}, we introduce observables in event-by-event fluctuation analysis.
In Sec.~\ref{subsec:CBWC} and \ref{subsec:VFC}, we explain the volume fluctuation and two correction methods called CBWC and VFC.
The centrality resolution effect is also explained with CBWC.
Then, in Sec.~\ref{sec:MCE}, multiplicity correlation is discussed as well as auto-correlation.
In Sec.~\ref{sec:NA}, we perform VFC and CBWC by using two models, toy model and UrQMD model.
Finally, in Sec.~\ref{sec:result} and in Sec.~\ref{sec:conclusion}, we discuss the results and summarize this paper.

\section{Observables}
\label{sec:Observables}
The moment and cumulant generating function, which are represented by $M(\theta)$ and $K(\theta)$, respectively, are defined as:
		\begin{eqnarray}
		\label{eq18_2}
		K(\theta)&=&{\rm ln}(M(\theta)),\\
    M(\theta)&=&\langle{e^{{\theta}N}}\rangle, 
		\end{eqnarray}
		where $N$ is the net-proton number and $\langle{X}\rangle$ represents the event-average of $X$.
		The $n^{th}$-order cumulants are defined by
		\begin{equation}
		\label{eq19}
		C_n=\frac{d^k}{{d\theta}^k}K(\theta)|_{\theta=0}.
		\end{equation}
		Up to the fourth-order cumulants can be written as:
		\begin{eqnarray}
		\label{eq20}
		C_1&=&\langle{N}\rangle,\\
		C_2&=&\langle{{(\delta{N})^2}}\rangle,\\
		C_3&=&\langle{{(\delta{N})^3}}\rangle,\\
		C_4&=&\langle{{(\delta{N})^4}}\rangle-3\langle{(\delta{N})^2}\rangle^2,
		\end{eqnarray}
		where $\delta{N}=N-\langle{N}\rangle$.

		In addition, $M$ (mean), $\sigma^2$ (variance), $S$ (skewness) and $\kappa$ (kurtosis) can be expressed by cumulants as:
		\begin{equation}
		\label{eq22}
		M=C_1,\;\;\;\;
		\sigma^2=C_2,\;\;\;\;
		S=\frac{C_3}{(C_2)^{\frac{3}{2}}},\;\;\;\;
		\kappa=\frac{C_4}{(C_2)^2}.
		\end{equation}
		By taking of different order of cumulants, the effect of the volume can be canceled out, which are expressed by moment products as:

		\begin{equation}
		\label{eq23}
		S\sigma=\frac{C_3}{C_2},\;\;\;\;\kappa\sigma^2=\frac{C_4}{C_2}.
		\end{equation}

\section{Volume Fluctuation and Multiplicity Correlation}
\label{sec:VF}
\subsection{Centrality Bin Width Correction (CBWC)}
\label{subsec:CBWC}
Theoretically, centralities are defined by the impact parameter $b$ or number of participants $N_{W}$.
However, these values cannot be measured directly in the real experiment even though recent study shows that the probability
distribution of $b$ in a given experiment-defined centrality can be reconstructed \cite{Rogly:2018ddx}.
Experimentally, centralities are determined using charged particle multiplicities.
For example, when we define 10\% centrality divisions, 0-10\% centrality corresponds to the largest 10\% multiplicity events.
However, even if we measure cumulants for each centrality bin, $N_{W}$ fluctuates event-by-event, which leads to the artificial enhancement of cumulants \cite{Broniowski:2017tjq}.
CBWC was proposed in order to suppress this VF.
Cumulants for each centrality bin are calculated by taking weighted average for each multiplicity bin as follows:
\begin{eqnarray}
\label{CBWC}
C_{n}=\sum_{r}{w_{r}C_{(n,r)}},~~~~
w_{r}=\frac{N_{r}}{\sum_{r}N_{r}},
\end{eqnarray}
where $N_{r}$ and $C_{(n,r)}$ are number of events and $n^{th}$-order cumulants in $r^{th}$ multiplicity bins, respectively.
From Eq.~(\ref{CBWC}), one may find that the result of CBWC directly depends on the centrality resolution.
Since we cannot measure all produced particles due to the limited experimental acceptance and detector efficiency,
we cannot eliminate VF completely even if cumulants are measured for each multiplicity bin \cite{VFC}.
We call this effect centrality resolution effect.

\subsection{Volume Fluctuation Correction (VFC)}
\label{subsec:VFC}
The VFC is derived from IPP \cite{VFC}, where measured net-particle was expressed by the sum of the net-particles from each source.
If we suppose $N_{W}$ is the number of sources, the moment generating function can be written as
\begin{eqnarray}
\label{eq_vfc0}
		M_{\Delta N}(\theta)= \left[ M _ { \Delta n } (\theta) \right] ^ { N _ { W } },
\end{eqnarray}
where $M_{\Delta N}(\theta)$ and $M_{\Delta n}(\theta)$ represent the moment generating function of measured net-particles and net-particles from each source, respectively.
Cumulants are given by the derivatives of the cumulant generating function $K_{\Delta N}(\theta)=\ln{(M_{\Delta N}(\theta))}$. 
Cumulants up to the fourth-order are given by \cite{VFC}:

\begin{eqnarray}
\label{eq_vfc1}
C_{1}{(\Delta{N})}&=& \langle{N_{W}}\rangle{C}_{1}{(\Delta{n})},\\
\label{eq_vfc2}
C_{2}{(\Delta{N})}&=&\langle{N_{W}}\rangle{C}_{2}{(\Delta{n})}+\langle{\Delta{n}}\rangle^2{C}_{2}{(N_{W})},\\
\label{eq_vfc3}
C_{3}{(\Delta{N})}&=&\langle{N_{W}}\rangle{C}_{3}{(\Delta{n})}+3\langle\Delta{n}\rangle{C}_{2}{(\Delta{n})}{C}_{2}{(N_{W})}\nonumber\\
                       &+&\langle{\Delta{n}}\rangle^3{C}_{3}{(N_{W})},\\
\label{eq_vfc4}
{C}_{4}{(\Delta{N})}&=&\langle{N_{W}}\rangle{C}_{4}{(\Delta{n})}+4\langle\Delta{n}\rangle{C}_{3}{(\Delta{n})}{C}_{2}{(N_{W})}\nonumber\\
&+&3{C}_{2}^{2}{(\Delta{n})}{C}_{2}{(N_{W})} +6\langle{\Delta{n}}\rangle^{2}{C}_{2}{(\Delta{n})}{C}_{3}{(N_{W})}\nonumber\\
&+&\langle{\Delta{n}}\rangle^4{C}_{4}{(N_{W})},
\end{eqnarray}
where ${C}_{n}{(\Delta{N})}$ and ${C}_{n}{(\Delta{n})}$ are the cumulants of $\Delta{N}$ and $\Delta{n}$ distributions, respectively.
The coefficients in Eq.~(\ref{eq_vfc2})-(\ref{eq_vfc4}) are also given by the Bell polynomials \cite{Skokov:2012ds}\cite{Broniowski:2017tjq}.
From Eq.~(\ref{eq_vfc2})-(\ref{eq_vfc4}), we find that $C_{n}(\Delta{N})$ is not only written by the sum of the $C_{n}(\Delta{n})$ but also $N_{W}$ cumulant ($C_{n}(N_{W})$) terms.
Those $C_n(N_{W})$ terms represent the VF background under the IPP assumption, which should be subtracted from measured cumulants.

\subsection{Multiplicity Correlation Effect}
\label{sec:MCE}
Measured cumulants are strongly affected not only by the centrality resolution effect which has already mentioned in Sec.~\ref{subsec:CBWC} but also by the auto-correlation effect.
When we use the same particles for both cumulant measurements and the centrality determinations, cumulants are artificially suppressed due to the auto-correlation effect \cite{Xiaofeng1}.
Experimentally, centralities are often determined in different kinematic window from cumulant measurements in order to avoid the auto-correlation.
In the case of net-proton analysis at the STAR experiment, centrality is determined in $|\eta|<1$, and net-proton is measured in $|y|<0.5$ \cite{net_proton}.
Therefore, if we use all charged particle multiplicities for centrality definitions, the same particles are used for both centrality definitions and cumulant measurements in $|\eta|<1$ and $|y|<0.5$, which leads to the auto-correlation.
In order to avoid the auto-correlation, centralities are determined by charged particle multiplicities excluding protons and anti-protons.
However, even though we use the different kinematic window or different particles, the positive multiplicity correlation between measured cumulants and the
 centrality may still exist.
This multiplicity correlation could also affect the measured cumulants.
We call this effect multiplicity correlation effect.
If we change the centrality definition, multiplicity correlation may also change, which affects the cumulants differently.
Therefore, it is more difficult to take multiplicity correlation into account than auto-correlation.
Multiplicity correlation effect will be studied by using UrQMD model in Sec.~\ref{sec:UrQMD}.

\onecolumngrid

\begin{figure}[H]   
\begin{center}   
\includegraphics[width=162mm]{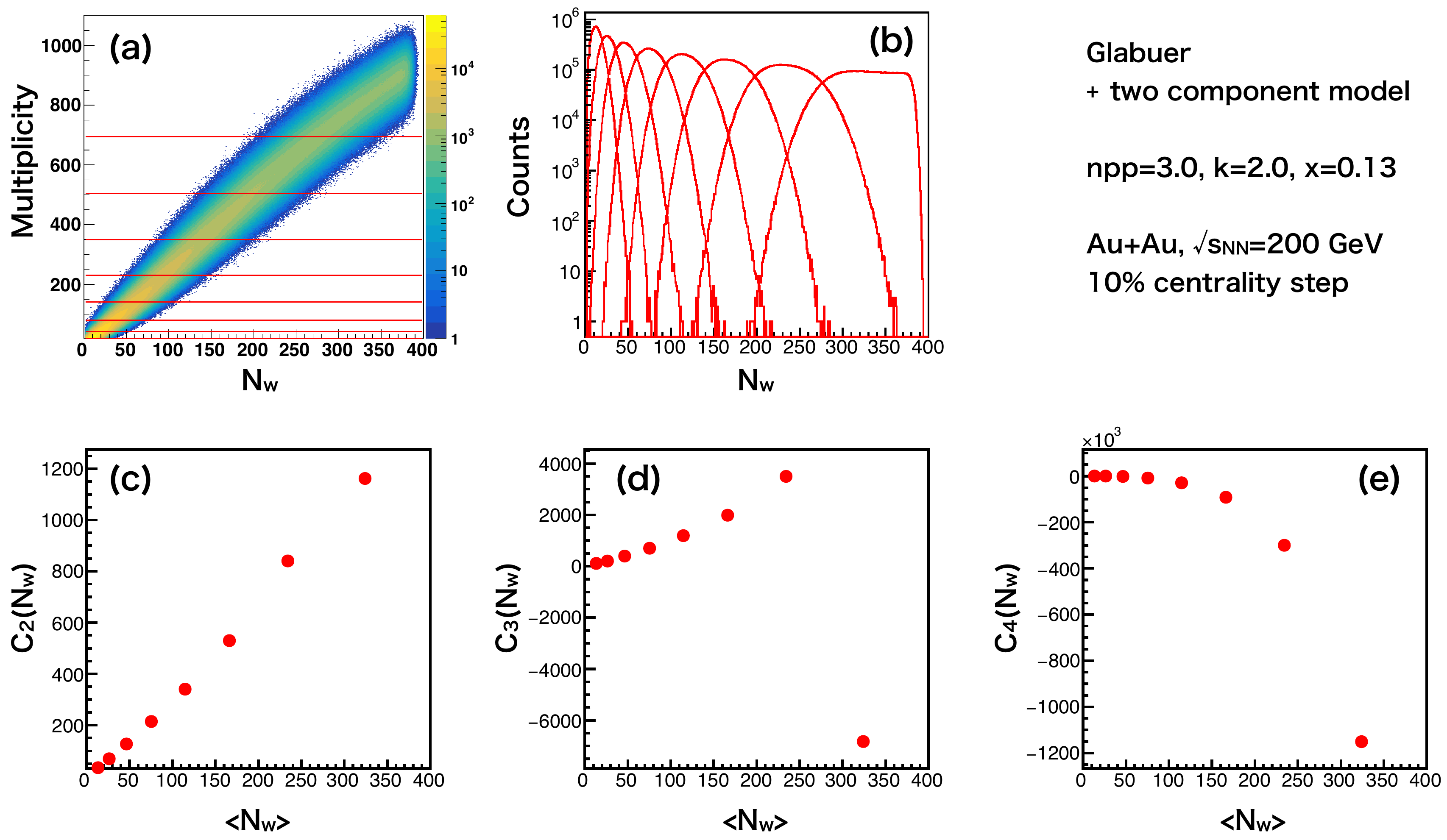}     
\caption{(a) Correlation between multiplicity and $N_{W}$ by Glauber simulation and two component model. (b) $N_{W}$ distributions for each centrality. (c)(d)(e) Second to the fourth-order $N_{W}$ cumulants as a function of $\langle{N_{W}}\rangle$. Number of events are 100 Million.}      
\label{fig:netp_Nwcumulant} 
\end{center}
\end{figure} 
\twocolumngrid

\section{Numerical Analysis}
\label{sec:NA}
In this section, we perform a similar numerical analysis in Ref.~\cite{VFC} in order to clarify the assumptions in VFC. In Sec.~\ref{sec:UrQMD}, UrQMD model is analyzed, which are compared to Sec.~\ref{sec:Toy} to point out the issue of VFC. The importance of the multiplicity correlation will be also discussed.

\subsection{Toy Model}
\label{sec:Toy}

In order to estimate the correction terms in the Eq.~(\ref{eq_vfc2})-(\ref{eq_vfc4}), we need to determine the cumulants of $N_W$ distribution. 
We use the Glauber model to define the centrality.
The final state multiplicity for centrality determination is produced from each source defined by two-component model \cite{Kharzeev},
\begin{equation}
		\label{eqsource}
N_{source}=(1-x)\frac{N_{W}}{2}+xN_{coll},
\end{equation}
where $N_{source}$ and $N_{coll}$ represent the number of source and number of collisions, respectively.
The Negative Binomial Distribution (NBD) is employed to implement the source-by-source multiplicity fluctuations.
The probability function of NBD is expressed by
		\begin{equation}
		\label{NBD8}
		P_{k,n_{pp}}(n)=\frac{\Gamma(n+k)}{\Gamma(n+1)\Gamma(k)}\left(\frac{n_{pp}}{ k } \right) ^ { n } \left( \frac { n_{pp} } { k } + 1 \right) ^ { - ( n + k ) },
		\end{equation}
where $n_{pp}$ represents the averaged number of produced particles from each source, and the $k$ controls the width of the NBD.
The $\Gamma$ represents the gamma fucntion and the $n$ is the natural number.
The parameters of Glauber simulation (Nucleon radius, differential cross section, etc.), the parameters of NBD ($n_{pp}, k$) and $x$ have been taken from the net-proton analysis from STAR in Au+Au collisions at \ene 200 GeV \cite{net_proton}, which are shown in the top right in Fig.~\ref{fig:netp_Nwcumulant}.  
Another parameter $\epsilon$ has been also introduced to describe multiplicity dependent efficiency of the detector.
Figure \ref{fig:netp_Nwcumulant}-(a) shows the correlation between multiplicity and $N_{W}$, where the centralities are defined by dividing the multiplicities into 10 classes, which are shown in red lines.
Figure \ref{fig:netp_Nwcumulant}-(b),(c),(d) and (e) show the $N_{W}$ distributions and the second to the fourth-order cumulants for each centrality.

Next, particles of interest, whose event-by-event distributions are analyzed, are generated from each participant nucleons independently (IPP) based on two Poisson distributions.
The parameters of the Poisson distributions ($\lambda_{p}$ and $\lambda_{\bar{p}}$) are defined as:
		\begin{eqnarray}
		\label{eqtoy}
		\lambda_{p}N_{W}&=&N_{p},\\
		\label{eqtoy2}
    \lambda_{\bar{p}}N_{W}&=&N_{\bar{p}}, 
		\end{eqnarray}
where we require that the efficiency corrected value of $C_1$ of protons ($N_{p}$) and anti-protons ($N_{\bar{p}}$) can be reproduced \cite{net_proton}, and
$N_{W}$ is given by the Glauber model event-by-event.
We can also use the fixed $N_{W}$ (averaged at each centrality) in order to see the pure cumulants without VF.


\subsection{UrQMD Model}
\label{sec:UrQMD}
\begin{figure}[htbp]   
\begin{center}   
\includegraphics[width=80mm]{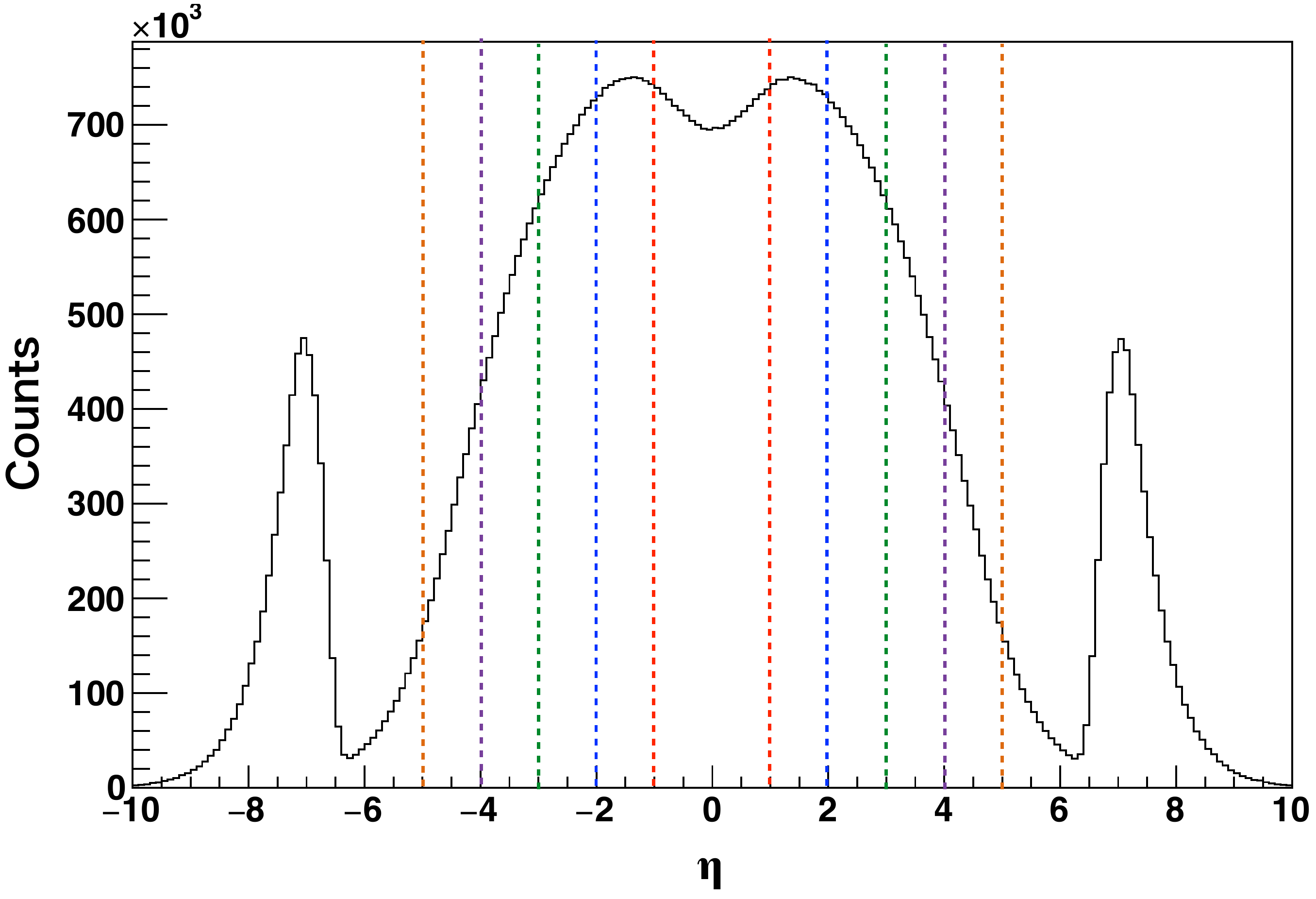}     
\caption{$\eta$ distribution of charged particles at \ene 200 GeV in UrQMD model. Dotted lines show $|\eta|=1$ ,2 ,3 ,4 ,5, respectively.}      
\label{fig:UrQMD_dist} 
\end{center}
\end{figure}

The Ultra relativistic Quantum Molecular Dynamics (UrQMD) \cite{UrQMD1998,UrQMD1999} is the microscopic transport model.
This model is based on hadron-hadron scattering, and can describe the excitation and decay of hadronic resonances and strings.
In this paper, the data of UrQMD model in Au+Au collisions at \ene 200 GeV is used.
The number of events are 45 Million.
Net-proton cumulants are measured in $|y|<0.5$, with the transverse momentum range $0.4<p_{T}<2.0$ GeV/$c$. The centrality is determined in $|{\eta}|<1$ excluding proton and anti-proton, which is the same as current net-proton cumulant analysis at the STAR experiment \cite{net_proton}.
We also define various centralities by using different kinematic window, $1<|{\eta}|<2$, $2<|{\eta}|<3$, $3<|{\eta}|<4$ and $4<|{\eta}|<5$ excluding protons and anti-protons.
Figure \ref{fig:UrQMD_dist} shows the $\eta$ distribution of charged multiplicity in UrQMD model, and each dotted line shows the boundary of the various centralities.
We can see two peaks structures around $|\eta|=7$-$8$, which correspond to the spectators.
In addition, we define centrality in $2<|{\eta}|<5$ by counting $\pi^{\pm}$ and $K^{\pm}$ multiplicities or $\pi^{\pm}$, $K^{\pm}$ and $p(\bar{p})$ multiplicities,
which are close to the EPD acceptance ($2.1<|{\eta}|<5.1$).
In the toy model in Sec.~\ref{sec:Toy}, true cumulants, which does not include VF, can be defined by using fixed $\langle{N_{W}}\rangle$ instead of ${N_{W}}$.
Since this method is not available in UrQMD, we introduce CBWC for each $N_{W}$ which we call "CBWC-N".
In UrQMD simulations, $N_{W}$ can be obtained directly.
Cumulants are calculated for each $N_{W}$ bin like a standard CBWC method shown in Eq.~(\ref{CBWC}).
In other words, bin-by-bin cumulants are measured in $N_{W}$ dimension in CBWC-N whereas cumulants are measured in multiplicity dimension in standard CBWC.
Since in VFC we define that the VF is arising from $N_{W}$ fluctuation, we can compute the pure cumulants using CBWC-N method.
We define two types of CBWC-N method.
The first definition, which we call "definition1", centralities are determined by charged particle multiplicities, and then cumulants are calculated for each $N_{W}$ bin.
The second definition, which we call "definition2", centralities are determined by dividing $N_{W}$ distribution, and then cumulants are calculated for each $N_{W}$ bin.
The CBWC-N of the first definition depends on how to determine centralities, such as $\eta$ region and centrality resolution, while the second
definition is only determined by $N_{W}$ distribution itself.
Cumulants are measured for each method and centrality definition, and compared to the results of toy model.
Regarding the VFC, there are two ways to calculate $N_{W}$ cumulants to get correction terms in Eq.~(\ref{eq_vfc2})-(\ref{eq_vfc4}).
One way is to perform Glauber fitting to the charged multiplicity distributions, divide it into centralities and extract $N_W$.
The other way is to calculate $N_{W}$ cumulants by $N_{W}$ distribution given by UrQMD model.
In order to apply VFC in more realistic situation, Glauber fitting is applied in this paper.
Centralities are also determined by the Glauber fitting to the multiplicity distributions.

\section{Results}
\label{sec:result}
\subsection{Toy Model Results}


\begin{figure*}[htbp]   
\begin{center}   
\includegraphics[width=170mm]{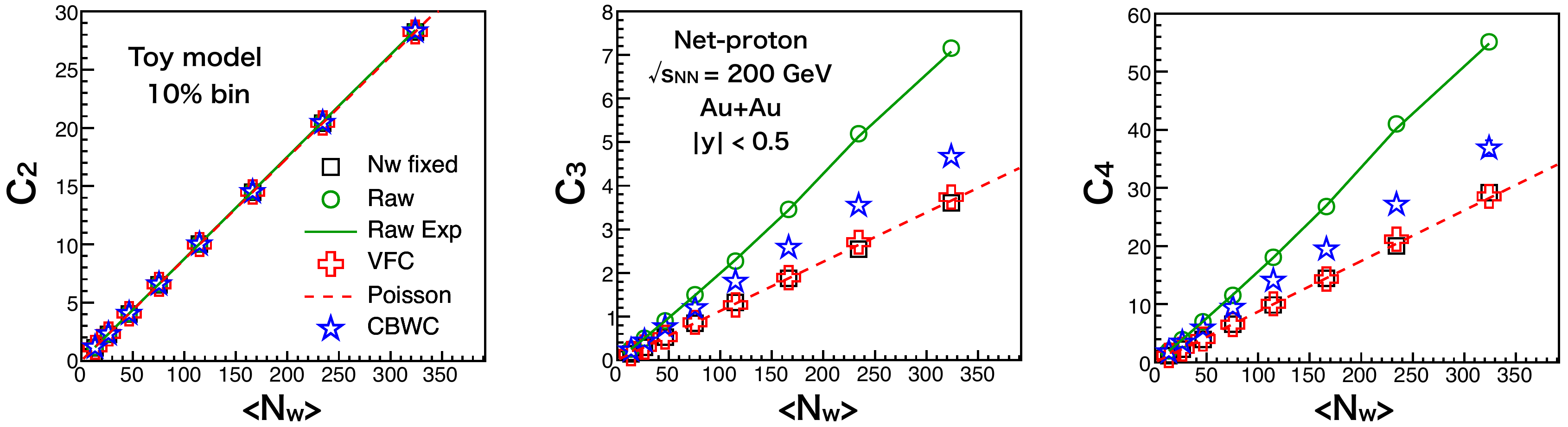}    
\caption{From the second to the fourth-order net-proton cumulants as a function of $\langle{N_{W}}\rangle$ by using toy model for 10\% centrality step.}      
\label{fig:toymodel_cumulant_netp} 
\end{center}
\end{figure*} 

%

\begin{figure*}[htbp]   
\begin{center}   
\includegraphics[width=150mm]{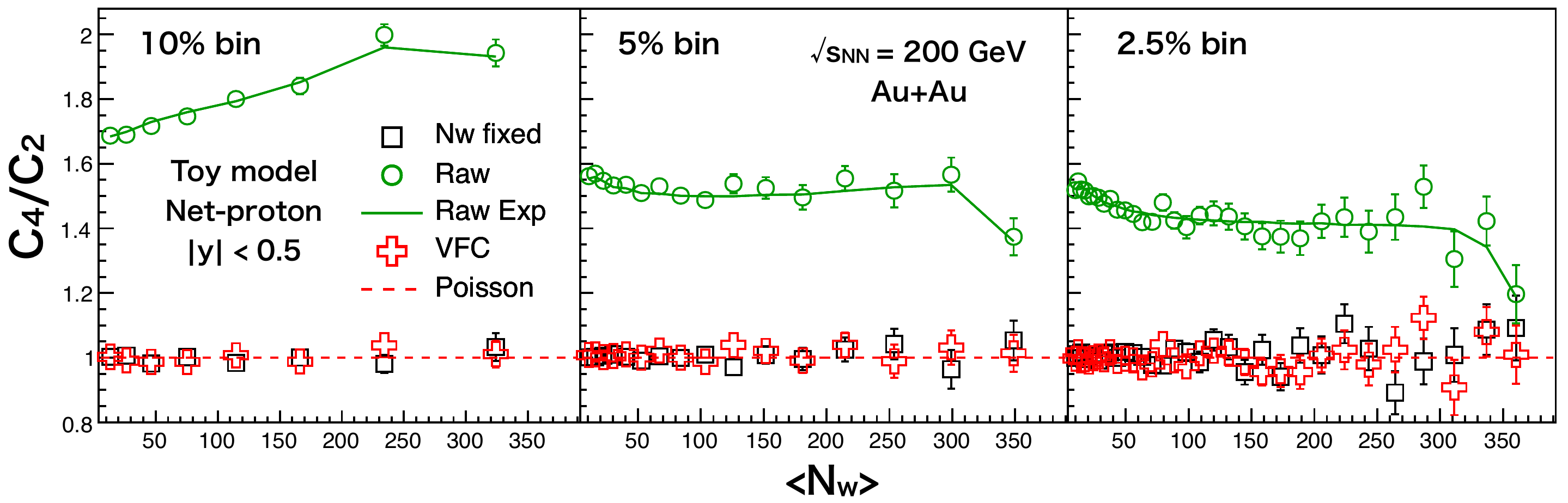}     
\end{center}
\caption{$C_{4}/C_{2}$ as a function of $\langle{N_{W}}\rangle$ by using toy model for 10\% (left), 5\% (middle) and 2.5\% (right) centrality step. The color and marker differences are the same as Fig.~\ref{fig:toymodel_cumulant_netp}.}      
\label{fig:toymodel_ratio_netp} 
\end{figure*} 


Figure \ref{fig:toymodel_cumulant_netp} shows the second to the fourth-order cumulants of net-proton distributions as a function of $\langle{N_{W}}\rangle$ with 10\% centrality step.
For black squares, $N_W$ is fixed at the value of the averaged number of participant nucleons ($\langle{N_{W}}\rangle$) in each centrality bin, they thus do not include VF. Green round symbols show the raw results without any corrections, which include the fluctuation of the $N_{W}$ in each centrality. 
$C_1$ is not shown because $C_{1}$ is not affected by VF. 
The red dotted lines and the green solid lines show the Poisson baselines and the expected values including $N_{W}$ fluctuation estimated from Eq.~(\ref{eq_vfc1})-(\ref{eq_vfc4}), respectively.
The $N_{W}$ fixed results are consistent with the Poisson baselines, and the raw results are also consistent with the expected baselines in all cases.
It is found that $C_{2}$ is unchanged for all cases.
This is because $\Delta{n}$ is small enough at \ene 200 GeV in Eq.~(\ref{eq_vfc2}). 
For $C_{3}$ and $C_{4}$, the raw results are larger than the $N_{W}$ fixed results, which indicates that the raw results are enhanced by VF.

Now let us compare the results of CBWC and VFC to see which method works better.
The VFC results (red cross) are consistent with the $N_{W}$ fixed results which indicates that VFC works well in this model.
As mentioned in Sec.\ref{subsec:VFC}, VFC relies on IPP assumption, and IPP is assumed in the toy model.
Therefore, it is natural that VFC works well in toy model by definition.
On the other hand, the CBWC results are smaller than the raw results but larger than the $N_{W}$ fixed results.
This indicates that CBWC can reduce VF but can not completely eliminate the VF.
Therefore, CBWC is not enough in toy model case. 
Figure \ref{fig:toymodel_ratio_netp} shows the $C_{4}/C_2$ ($=\kappa\sigma^2$) of net-proton distributions as a function of $\langle{N_{W}}\rangle$ for 10\%, 5\% and 2.5\% 
centrality divisions.
In 10\% centrality step, the raw results contain larger VF compared to the results with 5\% and 2.5\% step centralities, but the raw results are larger than VFC and the $N_{W}$ fixed results due to the risidual VF in any cases.
On the other hand, the VFC results are consistent with the $N_{W}$ fixed results in any cases and do not depend on centrality bin width by definition.

\subsection{UrQMD Results}
\begin{figure*}[htbp]   
\begin{center}   
\includegraphics[width=165mm]{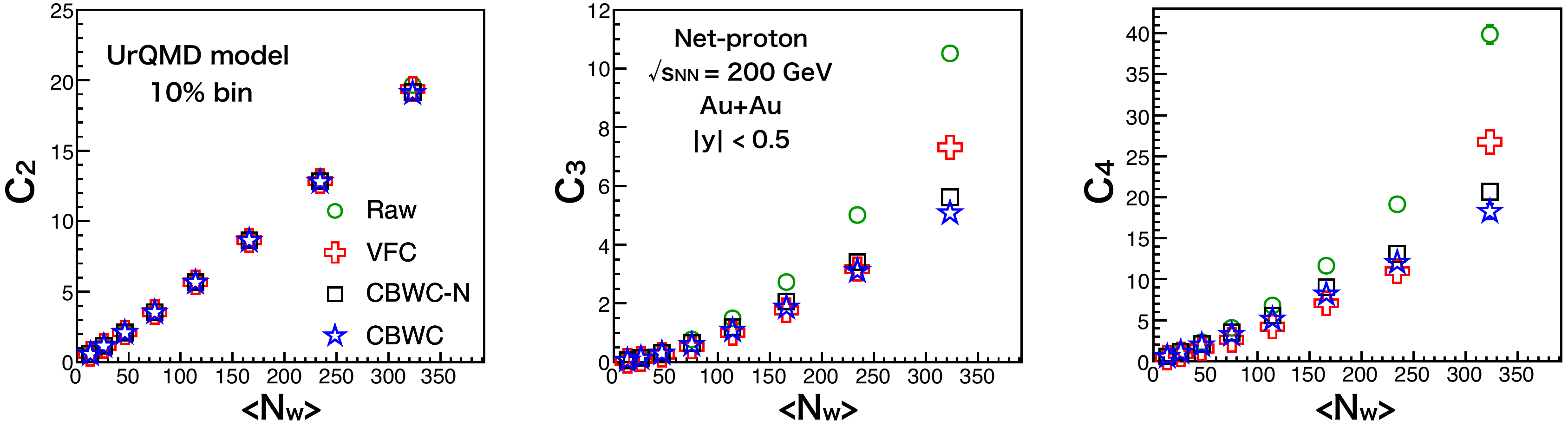}    
\caption{From the second to the fourth-order net-proton cumulants as a function of $\langle{N_{W}}\rangle$ by using UrQMD model for 10\% centrality step.}      
\label{fig:netp_UrQMD_cumulants} 
\end{center}
\end{figure*}

Figure \ref{fig:netp_UrQMD_cumulants} shows the second to the fourth-order cumulants of net-proton distributions as a function of $\langle{N_{W}}\rangle$ in UrQMD model for 10\% centrality step. 
Centralities are determined by counting $\pi^{\pm}$ and $K^{\pm}$ with the pseudo-rapidity range $|{\eta}|<1$.
The black squares are CBWC-N results, and the standard CBWC results are shown as the blue open stars.
Green round and red cross symbols represent the raw and VFC results, respectively.
$C_{2}$ is not affected by VF due to the small value of $\Delta{n}$, which is the same as Fig.~\ref{fig:toymodel_cumulant_netp}, but it can be found that results of $C_3$ and $C_4$ seem qualitatively different from Fig.~\ref{fig:toymodel_cumulant_netp}.
In Fig.~\ref{fig:toymodel_cumulant_netp}, the results of $N_{W}$ fixed, which correspond to CBWC-N in Fig.~\ref{fig:netp_UrQMD_cumulants}, are smaller than the CBWC results, and the VFC results are consistent with the $N_{W}$ fixed results. 
However, In Fig.~\ref{fig:netp_UrQMD_cumulants}, it is found that the results of CBWC-N are larger than standard CBWC results, and the VFC results are smaller than both of them except for the most central collisions, which cannot be explained by VF.
One of the reason is that IPP would be broken in UrQMD model.
As was explained in Sec.~\ref{subsec:VFC}, VFC formulas rely on the IPP assumption. So once it is broken, VFC loses its validity, which could be one of the reasons why VFC
does not seem to work well in UrQMD model.
However, it is still not clear why the CBWC results are smaller than the CBWC-N results at $C_3$ and $C_4$ in all centralities, because the CBWC results including risidual VF should be larger than the CBWC-N results which do not include VF.
In order to study these behaviour, we measure net-proton for various centrality definitions.


\begin{figure*}[htbp]   
\begin{center}   
\includegraphics[width=165mm]{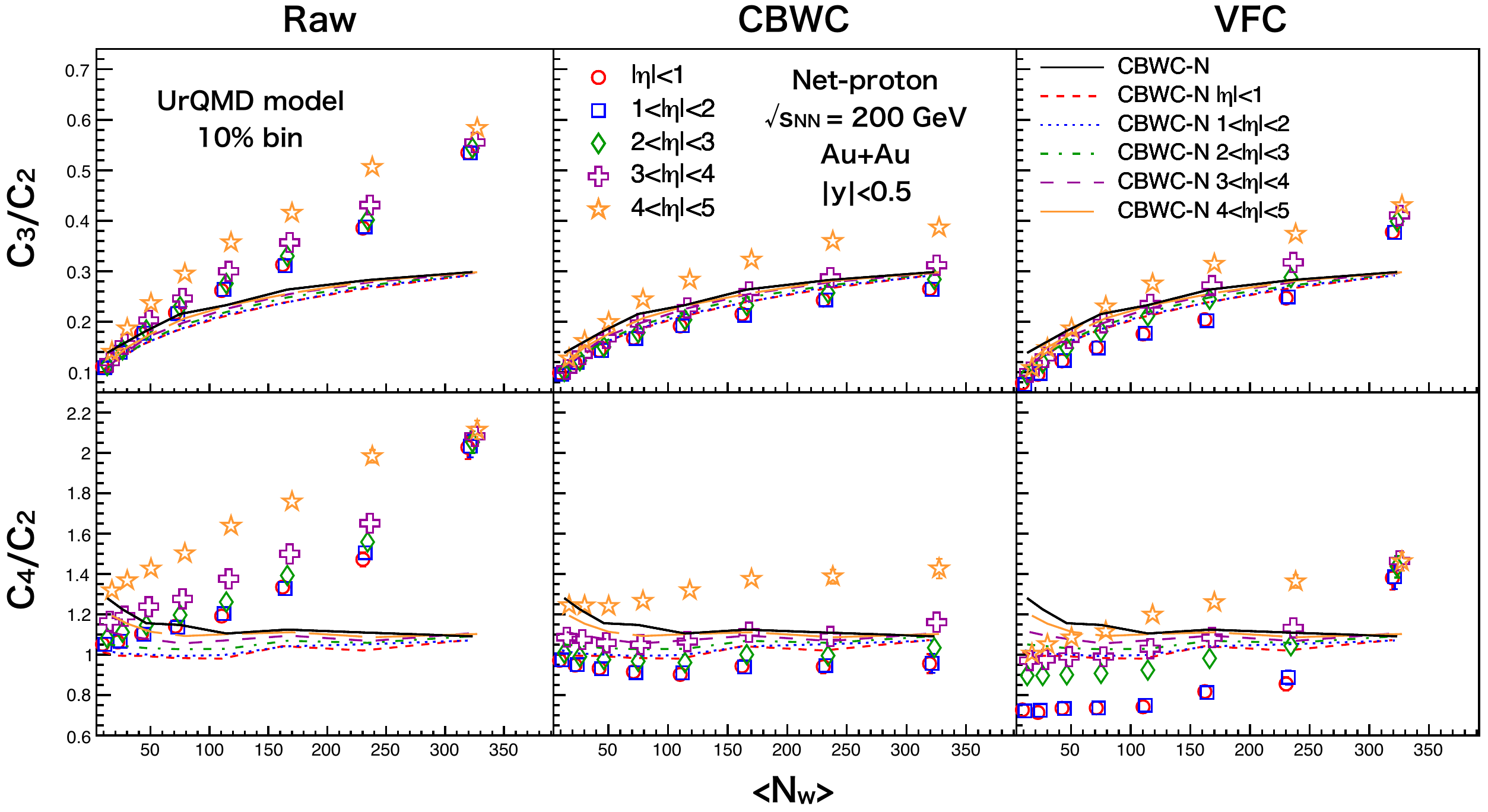}     
\caption{$C_{3}/C_{2}$ and $C_{4}/C_{2}$ of net-proton distributions as a function of $\langle{N_{W}}\rangle$ by using UrQMD model simulation for 10\% centrality divisions for different centrality definitions drawn in different markers. Centralities are determined in $|{\eta}|<1$, $1<|{\eta}|<2$, $2<|{\eta}|<3$, $3<|{\eta}|<4$ and $4<|{\eta}|<5$ excluding proton (anti-proton) drawn in different colors. Raw, CBWC and VFC results are shown from left to right. CBWC-N results by definition2 and definition1 are shown in black solid lines and colored dotted lines, respectively.}      
\label{fig:cent_dif_proton1} 
\end{center}
\end{figure*} 

%
\begin{figure*}[htbp]   
\begin{center}   
\includegraphics[width=155mm]{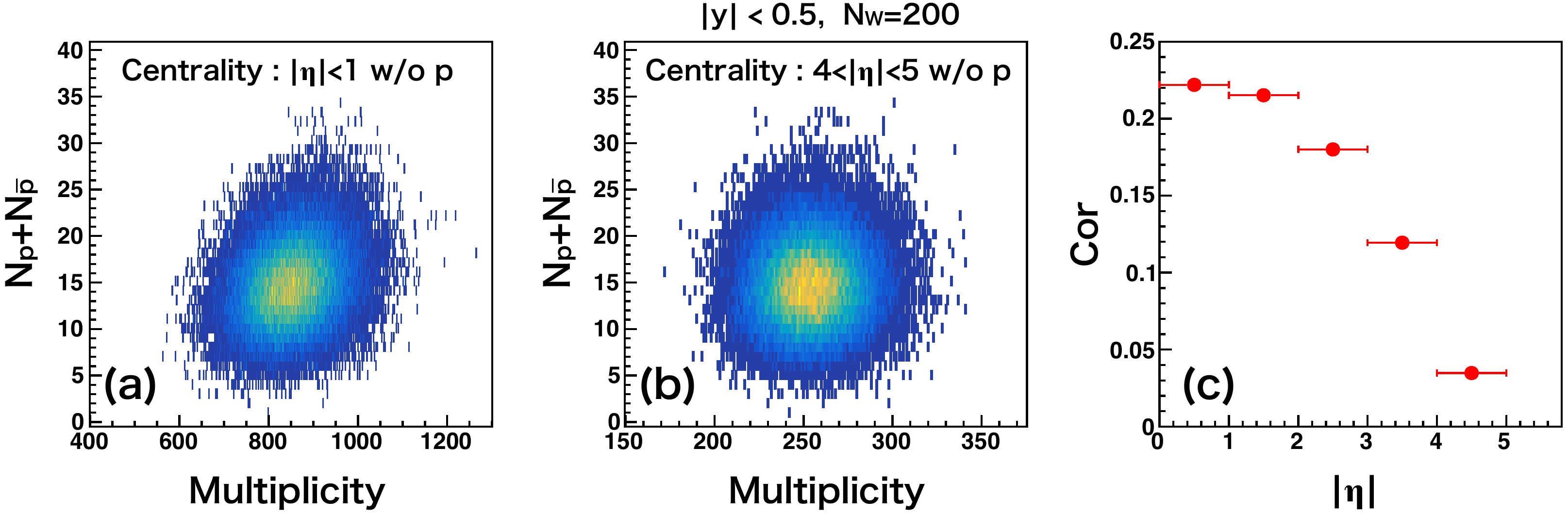}    
\caption{(a)(b) Correlations between multiplicities used for the centrality definitions and the number of protons and anti-protons in $|y|<0.5$ in UrQMD model. $N_{W}$ is fixed to 200, and centralities are defined within $|\eta|<1$ in (a) and $4<|\eta|<5$ in (b) excluding protons and anti-protons. (c) Correlation coefficient as a function of pseudo-rapidity regions used for the centrality definitions.}      
\label{fig:Corr} 
\end{center}
\end{figure*} 

Figure \ref{fig:cent_dif_proton1} shows $C_{3}/C_{2}$ ($=S\sigma$) and $C_{4}/C_{2}$ ($=\kappa\sigma^2$) of net-proton distributions as a function of centralities in UrQMD model for different centrality definitions and correction methods. 
Centralities are determined in $|{\eta}|<1$, $1<|{\eta}|<2$, $2<|{\eta}|<3$, $3<|{\eta}|<4$ and $4<|{\eta}|<5$ using charged pion and kaon multiplicities. 
Black solid lines show the CBWC-N results with "definition2", and colored dotted lines show the CBWC-N results with "definition1" for each centrality definition (See Sec.~\ref{sec:UrQMD}).
Red markers and lines correspond to the conventional centrality definitions.
In Fig.~\ref{fig:cent_dif_proton1}, the third and the fourth-order net-proton cumulants become larger with more forward centrality definitions for all cases.
There are two reasons to understand these differences.
The first possible reason is centrality resolution.
From Fig.~\ref{fig:UrQMD_dist}, it is obvious that multiplicities are smaller in forward pseudo-rapidity than mid-rapidity. Since the centrality resolution depends on the multiplicity, cumulants can be enhanced with the centrality definition in forward-rapidity due to the worse centrality resolution.
The second possible reason is a multiplicity correlation.
Figure \ref{fig:Corr}-(a) and (b) show the correlation between $N_{p}+N_{\bar{p}}$ measured in $|y|<0.5$ and the multiplicities used for the centrality determinations, where $N_{p}$ and $N_{\bar{p}}$ represent the number of protons and anti-protons, respectively, and $N_{W}$ is fixed to 200.
Centralities are determined within $|\eta|<1$ in Fig.~\ref{fig:Corr}-(a) and $4<|\eta|<5$ in Fig.~\ref{fig:Corr}-(b), and the correlation in Fig.~\ref{fig:Corr}-(a) seems stronger than Fig.~\ref{fig:Corr}-(b).
In order to discuss these multiplicity correlations more quantitatively, the correlation coefficient is introduced by,

		\begin{equation}
		\label{eqcorr}
		Cor=\frac{\sigma_{XY}}{\sigma_{X}\sigma_{Y}},
		\end{equation}
where $\sigma_{X}^{2}$ and $\sigma_{XY}$ denote the variance of $X$ and the covariance of $X$ and $Y$, respectively.
The values of $0<Cor<1$ indicate the positive correlations, and $-1<Cor<0$ indicate the negative correlations.
If there is no correlation between $X$ and $Y$, $Cor$ takes 0.
Figure \ref{fig:Corr}-(c) shows the correlation coefficient as a function of pseudo-rapidity window used for the centrality definitions.
From Fig.~\ref{fig:Corr}, the correlation become weaker when we define the centralities in forward pseudo-rapidity regions.
Therefore, cumulants become smaller in stronger multiplicity correlations, which can be deduced by comparing Fig.~\ref{fig:cent_dif_proton1} and Fig.~\ref{fig:Corr}.
Thus, as well as bad centrality resolution \cite{Xiaofeng1}, less multiplicity correlation may enhance the cumulants.
Next, let us discuss the results of CBWC-N in Fig.~\ref{fig:cent_dif_proton1}.
The CBWC-N results with definition2 (black solid lines) are larger than that of definition1 (colored dotted lines), and the results with definition1 for forward centrality definition are larger than that for mid-rapidity centrality definition.
These differences can also be explained by the multiplicity correlation.
CBWC-N with definition2 does not depend on the multiplicity correlation because the centrality is defined by $N_{W}$ itself.
On the other hand, in definition1, centralities are determined by charged multiplicities.
In CBWC in mid-rapidity centrality definitions, the CBWC results are smaller than the CBWC-N, which is also seen in Fig.~\ref{fig:netp_UrQMD_cumulants}.
However, when we define centralities in forward pseudo-rapidity regions, the CBWC results are larger than CBWC-N.

\begin{figure}[htbp]   
\begin{center}   
\includegraphics[width=168mm]{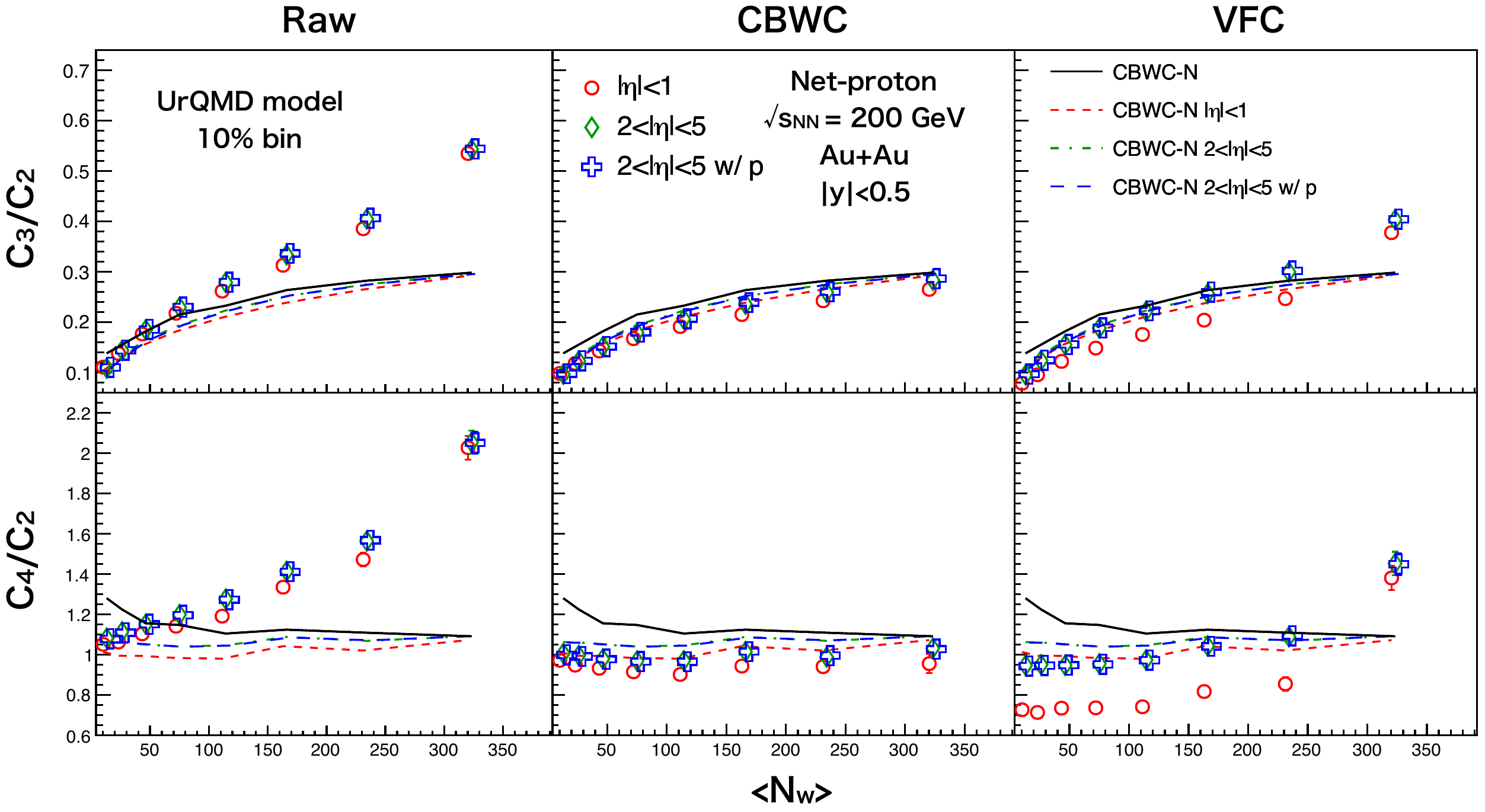}     
\caption{$C_{3}/C_{2}$ and $C_{4}/C_{2}$ of net-proton distributions as a function of $\langle{N_{W}}\rangle$ by using UrQMD model simulation for 10\% centrality divisions for different centrality definitions drawn in different markers. Centralities are determined in $|{\eta}|<1$, $2<|{\eta}|<5$ excluding proton (anti-proton) and $2<|{\eta}|<5$ including proton (anti-proton) drawn in different colors. Raw, CBWC and VFC results are shown from left to right. CBWC-N results by definition2 and definition1 are shown in black solid lines and colored dotted lines, respectively.}      
\label{fig:cent_dif_proton2} 
\end{center}
\end{figure}

We have been discussed that bad centrality resolutions and less multiplicity correlations may enhance the cumulants in forward rapidity regions.
However, it is not clear which is the dominant reason of the enhancements in forward rapidity regions, bad centrality resolutions or less multiplicity correlations.
In order to answer this question, we define centralities with good resolution in forward rapidity regions.
Figure \ref{fig:cent_dif_proton2} is almost the same as Fig.~\ref{fig:cent_dif_proton1} but only centrality definitions are different. 
Color differences represent the different centrality definitions, the pseudo-rapidity range with $|{\eta}|<1$ and $2<|{\eta}|<5$.
Centralities are determined by $\pi^{\pm}$ and $K^{\pm}$.  
For $2<|{\eta}|<5$, we also try including protons and anti-protons, which assumes EPD in the STAR experiment.
The difference between with and without proton (anti-proton) for centrality definition is small.
Furthermore, these systematic differences for various centrality definitions of VFC are larger than CBWC, and the differences between VFC and CBWC-N are larger than CBWC and CBWC-N.
These results imply CBWC might be better than VFC in UrQMD model, while VFC was the best method in the toy model by definition.
If we apply VFC to the experimental data, we have to note that IPP is expected to be broken in the real experiment.

\section{Conclusions}
\label{sec:conclusion}
Validity of the volume fluctuation correction and importance of the multiplicity correlation on higher cumulants are discussed by using toy model assuming IPP and UrQMD simulation.
In toy model, even if CBWC has been applied or using 2.5\% centrality division, effect from VF can not be removed completely while VFC works well in all cases by definition.
In UrQMD, however, VFC does not seem to work well, which would be because IPP model is broken.
In addition, we found that cumulants depend on how to determine centralities because of centrality resolution and multiplicity correlation effect.
The multiplicity correlation seems enhance the measured cumulants, which might indicate that cumulants measured by EPD centrality definitions may be larger than
that in mid-rapidity. 
We have to consider these effects for future higher-order cumulants analysis.
In addition, these systematic differences for various centrality definitions of VFC are larger than CBWC, and the CBWC results are closer to the CBWC-N results than VFC.
Therefore, we have to consider these effects if VFC is applied to experimental data.
In this paper, we does not propose new correction methods or revolutionary solutions.
However, we find various effects which should be considered when we apply CBWC and VFC to the experimental data.
Therefore, both results should be compared in the real experiment.

\section{Acknowledgement}
We thank Ito Science Foundation (2017), JSPS KAKENHI Grant Number 25105504 and International travel support for graduate student in University of Tsukuba.
We also acknowledge support from the MoST of China 973-Project No. 2015CB856901, the National Natural Science Foundation of China under Grants (No. 11575069, 11828501, 11890711 and 11861131009), Fundamental Research Funds for the Central Universities NO. CCNU19QN054 and China Postdoctoral Science Foundation funded project 2018M642878.

\onecolumngrid

\twocolumngrid

\vspace{2pt}

\bibliography{main}
\end{document}